\documentclass[aps,pra,preprint]{revtex4-2}

\usepackage{hyperref}
\usepackage{amssymb,amsmath}
\usepackage{graphicx}

\begin{document}

\title{Nonlinear dynamics of molecular superrotors}

\date{March 1, 2023}

\author{C. \surname{Chandre}}
\email{cristel.chandre@cnrs.fr}
\affiliation{CNRS, Aix Marseille Univ, I2M, 13009 Marseille, France}
  
\author{J. Pablo \surname{Salas}}
\email{josepablo.salas@unirioja.es}
\affiliation{\'Area de F\'{\i}sica, Universidad de la Rioja, 26006 Logro\~no, La Rioja, Spain}

\begin{abstract}
We consider a diatomic molecule driven by a linearly polarized laser pulse with a polarization axis rotating with a constant acceleration. This setup is referred to as optical centrifuge, and it is known to lead to high-angular momenta for the molecule (superrotor states) and, possibly, to dissociation. Here we elucidate the dynamical mechanisms behind the creation of superrotor states and their dissociation. We unravel the role of the various parameters of the laser field in these processes by considering reduced Hamiltonian models encapsulating the different phases in the creation of superrotor states,  possibly leading to dissociation.
\end{abstract}

\maketitle 

\section*{Introduction}

The interaction between matter and laser light has played a major role in probing matter at unprecedented temporal and spatial scales, providing more complex dynamical processes than anticipated. For example, when an atom is subjected to an intense laser field, the complex interplay between the electron-core Coulomb force and the force exerted by the electric field leads to single to multiple electron ionizations (see, e.g., Refs.~\cite{Becker2008,Becker2012} and references therein), evidencing the pivotal role of electron-electron interaction in strong-field processes. At the same time, the identification of the dynamical processes resulting from these interactions has opened up a broad and multi-disciplinary research field to the manipulation of matter by fine tuning the parameters of the laser. For instance, when a molecule is subjected to an optical wave, the interaction between the electric field of the wave and the induced dipole moment of the molecule, makes it possible to control its rotation (see, e.g., Refs.~\cite{Ohshima2010,MacPhail-Bartley2020}) and its spatial orientation or alignment (see, e.g., Refs.~\cite{Larsen2000,Peronne2003,Stapelfeldt2003,Poulsen2004,Daems2005,Hertz2007}). In particular, molecular alignment is of great importance in chemical reactions since the initial relative orientation between reactants has, in many situations, a great impact on the reaction rate~\cite{Xie2014}.

Another example of the manipulation of matter using lasers is the optical centrifuge for molecules. 
This technique was proposed in Ref.~\cite{Karczmarek1999} to control and bring molecules to extreme rotational states. In a nutshell, an optical centrifuge consists of an infrared linearly polarized laser pulse, whose polarization axis rotates with constant angular acceleration. Because a molecule in the presence of that field will tend to align along the polarization axis of the laser, the molecule will be forced to follow the laser rotation as well. Therefore, molecules in an optical centrifuge can be excited to such very high rotational levels that they eventually dissociate \cite{Villeneuve2000,Spanner2001a,Spanner2001b,Hasbani2002}. Ultimately, the ability of the optical centrifuges to create those so-called superrotor states, has been widely used in a large number of experiments in, for example, molecular spectroscopy~\cite{Yuan2011a,Korobenko2014b, Milner2017a}, molecular dynamics~\cite{Toro2013,Korobenko2014a,Milner2014,Murray2018,Korobenko2015a,Milner2015a,Korobenko2016, Korobenko2015b, Milner2017b} or to study molecular magnetic properties~\cite{Milner2015b,Faucher2016,Floss2018}.

The interaction in an optical centrifuge for molecules is of nonlinear nature. To a large extent, this interaction is characterized by the presence of a Coriolis term in the Hamiltonian that appears explicitly when the problem is formulated in a reference frame rotating with the polarization axis of the laser used in the centrifuge. Parameters of the laser such as the laser pulse profile, the angular acceleration of the polarization axis of the laser and the strength of its electric field are involved in the dynamical processes and can be used to control the amount and type of superrotor states. In order to differentiate the role played by each of these parameters in the creation of the molecular superrotors, several investigations have considered a classical mechanical treatment~\cite{Karczmarek1999,Spanner2001b,Hasbani2002}. In these investigations, different theoretical approaches are proposed, all of them revealing the essential role the Coriolis term plays in the high-angular acceleration of molecules in an optical centrifuge. 

Here the main goal of our article is unravel the role of each parameter of the laser in the building up of superrotor states. We also would like to understand why some superrotor states end up dissociating and some others do not. By using nonlinear dynamics, we are able to precisely identify the dynamical mechanisms responsible for the creation and dissociation of these superrotor states. By identifying and analyzing reduced Hamiltonian models we are able to assess the role each parameter of the laser plays in these processes. More precisely, following a similar scheme to the one we used in Refs.~\cite{Chandre2017,Chandre2019}, we study the classical dynamics of a diatomic molecule in an optical centrifuge. As a model example, we take the Cl$_2$ molecule which has been used in Refs.~\cite{Karczmarek1999,Spanner2001b}. Besides the kinetic terms and the potential energy between the Cl atoms, the rovibrational Hamiltonian of the system includes the interaction between the molecular polarizability and the laser pulse. The resulting Hamiltonian model for the dynamics of the molecule in an optical centrifuge has 3 + 1/2 degrees of freedom (i.e., the 3 degrees of freedom of the molecule plus the explicit time dependence of the laser field). It should be noted that the explicit time dependence is twofold: On the one hand, we have the polarization axis of the laser rotating with constant angular acceleration. On the other hand, we have an additional time dependence coming from the laser pulse envelope, which consists of a ramp-up, a plateau and a ramp-down. None of these two explicit time dependencies are periodic. 

The questions we address here are: Why do some molecular states lead to superrotors and some even nearby states do not? Why do some superrotor states end up dissociating and some others do not? What are the roles of the laser parameters in the nonlinear dynamics of molecular superrotors?   

In Sec.~\ref{sec:model}, we explicit the model we use for the interaction between the laser pulse and the Cl$_2$ molecule. In particular, we briefly review the method to average the dynamics over one laser period. We also construct the polarizabilities of the molecule, a crucial ingredient in the model. In Sec.~\ref{sec:proba}, we detail the computation of dissociation probabilities and highlight some puzzling characteristics when the amplitude of the laser field is varied. We conclude this section with a list of questions regarding the dynamical mechanisms behind these probability curves. In Sec.~\ref{sec:analysis}, we analyze two reduced Hamiltonian models to fully characterize these dynamical mechanisms: one model to characterize the mechanisms for the creation of superrotor states, the other one for the possible dissociation of these states.  

\section{\label{sec:model} Optical Centrifuge for Diatomic Molecules: The Hamiltonian model}

\subsection{The interaction Hamiltonian}

Under Born-Oppenheimer approximation, we consider a diatomic molecule in the presence of a strong linearly polarized laser field ${\bf E}(t)$ of amplitude $F_0$, frequency $\omega$ and pulse profile $f(t)$.
If the polarization axis of the laser is slowly rotating in the $xy$-plane, the electric field ${\bf E}(t)$ writes as \cite{Karczmarek1999}
\begin{equation}
\label{laser}
{\bf E}(t) = F_0 f(t) [{\bf \widehat x} \cos \Phi(t) + {\bf \widehat y} \sin\Phi(t)] \cos \omega t,
\end{equation}
\noindent
where $\Phi(t)$ is the instantaneous polarization angle of the laser. We consider the situation where the polarization axis rotates with constant angular acceleration $\beta$, so that $\Phi(t)=\beta  t^2/2$.
Hereafter, we assume that the laser pulse profile $f(t)$ is a sine ramp-up function
$$
\label{pulse}
f(t)=\left \{
\begin{array}{ccc} 
\sin\left(\frac{\pi t}{2 t_{\rm u}}\right) & \mbox{if} & 0 \le t < t_{\rm u}, \\
1 & \mbox{if} & t_{\rm u} \le t \le t_{\rm u}+t_{\rm p}, \\
\sin\left(\frac{\pi (t_{\rm u}+t_{\rm p}+t_{\rm d} - t)}{2 t_{\rm d}}\right) & \mbox{if} & \quad t_{\rm u}+t_{\rm p}< t \le t_{\rm u}+t_{\rm p}+t_{\rm d}, \\
0 & \mbox{otherwise} , &
\end{array}
\right.
$$
where $t_{\rm u}$ is the duration of the ramp-up of the laser field, $t_{\rm p}$ the duration of its plateau and $t_{\rm d}$ the duration of the ramp-down.
This sine ramp-up envelope has been previously used in Refs.~\cite{Dion1999,Chandre2017}, and it is used to mimic experimental laser pulses with a rather smooth ramp-up and ramp-down~\cite{Trippel2014}.
With the electric field \eqref{laser} we irradiate a diatomic molecule. In Cartesian coordinates $(x, y, z)$, the dynamics resulting from the interaction between the laser field and the diatomic molecule is governed by the Hamiltonian
\begin{equation}
\label{ham0}
{\cal H} = \frac{p_x^2+p_y^2+p_z^2}{2\mu} +\varepsilon(r) -{\bf d}(r)\cdot {\bf E}(t)+ {\bf E}(t) \cdot \alpha(x,y,z) {\bf E}(t),
\end{equation}
where $r$ is the interatomic distance $r =(x^2+y^2+z^2)^{1/2}$ and $\alpha$ is the polarizability matrix of the diatomic molecule. Here $\mu$ is the reduced mass of the diatomic molecule and $\varepsilon(r)$ is its potential energy curve.

If the frequency $\omega$ of the laser is much larger than the rotational frequencies of the diatomic molecule (typically, $\omega$ is in the infrared regime), Hamiltonian \eqref{ham0} can be averaged over one laser period (e.g., using a canonical Lie transform). The averaged Hamiltonian becomes
\begin{equation}
\label{ham1}
{\cal H} = \frac{p_x^2+p_y^2+p_z^2}{2 \mu} + \varepsilon(r) + V_L(x,y,z,t),
\end{equation}
where the interaction potential $V_L$ of the molecule with the laser is given by
$$
V_L(x,y,z,t)= -\frac{1}{4} \ F_0^2 \ f(t)^2 [\Delta\alpha(r) \cos^2\theta_l + \alpha_\bot(r)].
$$
In the above expression $\Delta\alpha(r)=\alpha_\parallel(r)-\alpha_\bot(r)$, with $\alpha_\parallel(r)$ and $\alpha_\bot(r)$ being the parallel and perpendicular components of the polarizability, and $\theta_l$ is the angle between the molecular axis and the electric field ${\bf E}(t)$.

When the problem is formulated in a reference frame $(x',y',z')$ rotating with the frequency $\Omega(t)=\dot \Phi(t)=\beta t$ of the polarization axis (by applying a canonical change of coordinates), the time dependence is reduced to $\Omega(t)$ and to the pulse profile $f(t)$, in such way that Hamiltonian \eqref{ham1} becomes
\begin{equation}
\label{ham2}
{\cal H'} = \frac{p_x'^2+p_y'^2+p_z'^2}{2 \mu} -\Omega(t)  (x'  p_y' -y' p_x')+\varepsilon(r) -\frac{F_0^2}{4}  f(t)^2 [\Delta\alpha(r) \cos^2\theta_l+\alpha_\bot(r)] .
\end{equation}
\noindent
In this rotating frame, the $x'$-axis marks the direction of the polarization vector of the electric field, and the angle $\theta_l$ satisfies $\cos \theta_l = x'/r$. After dropping primes on coordinates and momenta for simplification, Hamiltonian \eqref{ham2} reads
\begin{equation}
\label{ham3}
{\cal H} = \frac{p_x^2+p_y^2+p_z^2}{2 \mu} -\Omega(t) (x  p_y -y p_x)+\varepsilon(r) -\frac{F_0^2}{4}  f(t)^2 \left[\Delta\alpha(r) \frac{x^2}{r^2}+\alpha_\bot(r)\right].
\end{equation}
It is convenient to formulate Hamiltonian~\eqref{ham3} in spherical (canonical) coordinates $(r, \theta, \phi, p_r, p_{\theta},  p_{\phi})$ where $\theta \in [0,\pi]$ and $\phi \in [-\pi, \pi[$.
In this coordinate system, Hamiltonian \eqref{ham3} read as
$$
{\cal H} = \frac{1}{2 \mu}\left(p_r^2+\frac{p_{\theta}^2}{r^2}+\frac{p_{\phi}^2}{r^2 \sin^2 \theta}\right) + \varepsilon(r) -\Omega(t) p_{\phi} -\frac{F_0^2}{4} f(t)^2 [\Delta \alpha(r) \sin^2\theta \cos^2\phi+\alpha_\bot(r)].
$$
We notice that the manifold defined by $\theta=\pi/2$ and $p_{\theta}=0$ is invariant under the dynamics. In this manifold, the system reduces to the following Hamiltonian system in the canonical coordinates $(r, \phi, p_r, p_{\phi})$,
\begin{equation}
\label{ham2D}
{\cal H}_{\rm 2D} = \frac{1}{2 \mu}\left(p_r^2+\frac{p_{\phi}^2}{r^2}\right) + \varepsilon(r) -\Omega(t) p_{\phi} -\frac{F_0^2}{4} f(t)^2 [\Delta \alpha(r) \cos^2\phi+\alpha_\bot(r)].
\end{equation}
In the manifold defined by $\theta=\pi/2$ and $p_{\theta}=0$, the molecular motion takes place in the $xy$-plane, so that $p_{\phi}$ is the total angular momentum of the molecule.
For the sake of simplicity, we reduce our study to that manifold $\theta=\pi/2$ and $p_{\theta}=0$.

\subsection{A case study: The chlorine molecule}

In order to study the dynamics arising from Hamiltonian \eqref{ham2D}, we consider the chlorine molecule Cl$_2$ ($\mu \approx 32548.53$ a.u.) as a model example.

\subsubsection{Potential energy curve}

The electronic potential energy curve $\varepsilon(r)$ for the chlorine molecule is modeled by means of a Morse potential
$$
\varepsilon(r)=D_{\rm e} \left[1-\exp (-\gamma (r-r_{\rm e}))\right]^2-D_{\rm e},
$$
where $r_{\rm e}\approx 3.7560$ a.u. is the equilibrium distance,  $D_{\rm e}\approx 0.0915$ a.u.  is the potential well depth, and $\gamma \approx 1.0755$ a.u. is the width parameter \cite{Hormain2015}.

\subsubsection{Parallel and perpendicular polarizabilities}

\begin{figure}
\includegraphics[scale=0.4]{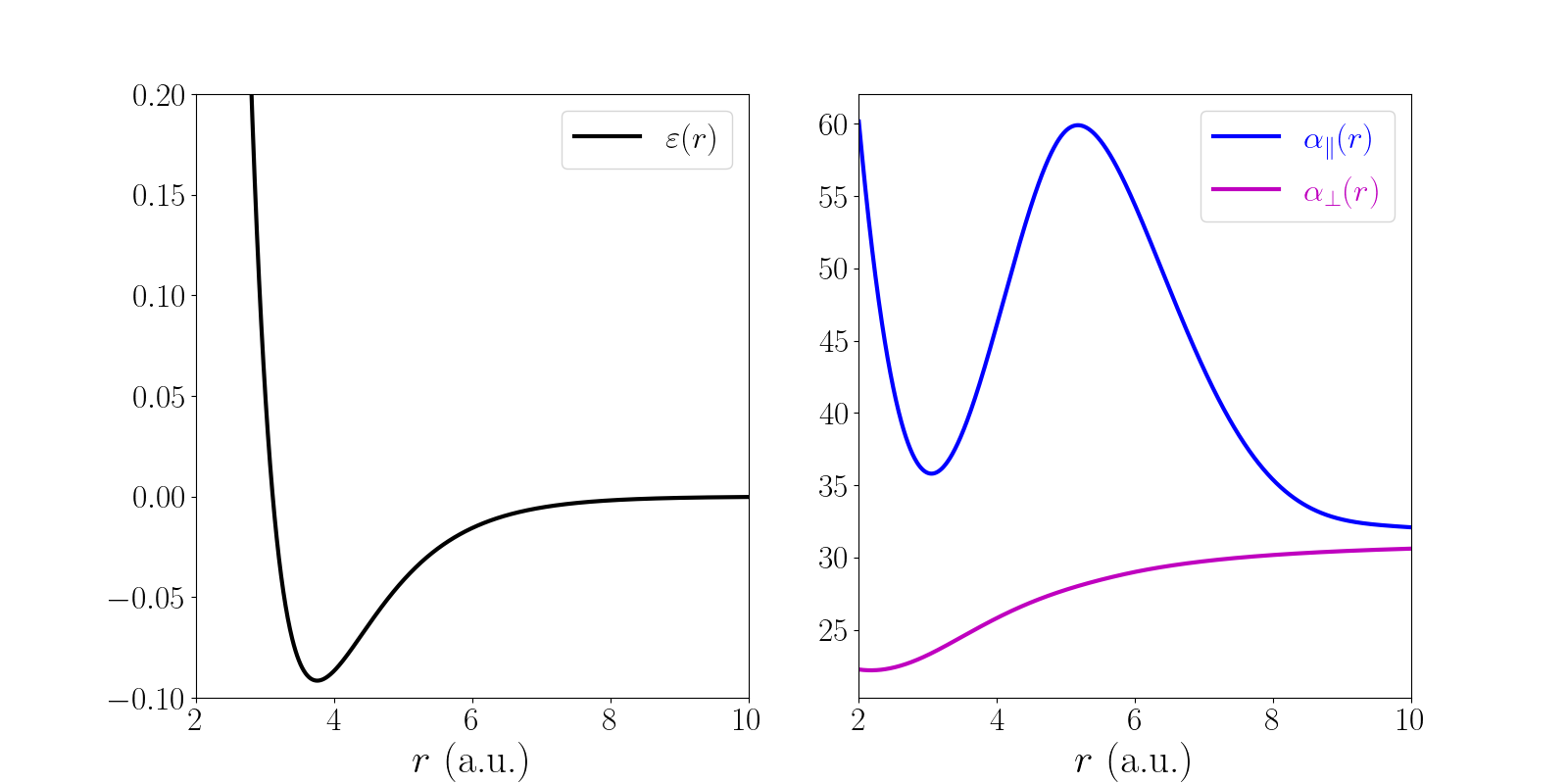}
\caption{\label{fi:curvas1} Electronic potential energy curve $\varepsilon(r)$ (left panel),
and parallel $\alpha_\parallel(r)$ and perpendicular $\alpha_\bot(r)$ components of the molecular polarizability (right panel) of the Cl$_2$ molecule. The values are in atomic units.}
\end{figure}

We follow Refs.~\cite{Maroulis1992,Maroulis2004} to construct analytic functions for $\alpha_\parallel(r)$ and $\alpha_\bot(r)$.
In the range of small $r$, the Cl$_2$-polarizability functions in atomic units are given by polynomials of the form
\begin{subequations}
\begin{eqnarray}
\label{para}
\alpha_\parallel^{\rm SR}(r) &=& 42.13 + 15.40(r - r_{\rm e}) + 5.40(r - r_{\rm e})^2 - 5.25(r - r_{\rm e})^3,\\
\label{per}
\alpha_\bot^{\rm SR}(r)&=&25.29 + 2.87(r - r_{\rm e}) - 0.09(r - r_{\rm e})^2 - 0.42(r - r_{\rm e})^3,
\end{eqnarray}
\end{subequations}
where $r$ and $r_{\rm e}$ are also in atomic units. 
On the other side, the asymptotic behavior of the polarizabilities is well described by the Silberstein expressions
\cite{Silberstein1917,Jensen2002}:
\begin{subequations}
\begin{eqnarray}
\label{silber1}
\alpha_\parallel^{\rm LR}(r) &=& \frac{\alpha_{\rm Cl_2} + 4 \alpha_{\rm Cl}^2/r^3}{1 - 4 \alpha_{\rm Cl}^2/r^6},\\
& & \nonumber \\
\label{silber2}
\alpha_\bot^{\rm LR}(r) &=& \frac{\alpha_{\rm Cl_2} - 2 \alpha_{\rm Cl}^2/r^3}{1 -  \alpha_{\rm Cl}^2/r^6},
\end{eqnarray}
\end{subequations}
where $\alpha_{\rm Cl}\approx 15.5421$ a.u. is the atomic polarizability of the Cl atom and $\alpha_{\rm Cl_2} =2 \alpha_{\rm Cl}$.
In the middle internuclear distances, we find $\alpha_{\parallel,\bot}$ joining the polarizability functions for small and large $r$ [given by Eqs. \eqref{para}-\eqref{per} and \eqref{silber1}-\eqref{silber2}, respectively]. As joining functions we use two polynomials of degree five,
\begin{eqnarray*}
\alpha_\parallel^{\rm MR}(r) &=& \sum_{k=0}^{5} a_k r^k,\\
\alpha_\bot^{\rm MR}(r) &=& \sum_{k=0}^{5} b_k r^k.
\end{eqnarray*}
The $a_k$ and $b_k$ coefficients are found imposing continuity conditions up to the second derivatives \cite{Maroulis2004}. The joining points were taken at $r_1=5$ a.u. and $r_2=10$ a.u. for $\alpha_\parallel(r)$, and at $r_1=3$ a.u. and $r_2=6$ a.u. for $\alpha_\bot(r)$. The values of the corresponding coefficients are given in Table~\ref{ta:tabla1}. In Fig.~\ref{fi:curvas1}, the potential energy curve $\varepsilon(r)$ and the fitted curves $\alpha_{\parallel,\bot}(r)$ are displayed.
\begin{table*}
\caption{\label{ta:tabla1} Values (in atomic units) of the parameters for the medium-range behavior of the polarizability curves $\alpha_{\parallel,\bot}(r)$.}
\begin{ruledtabular}
\begin{tabular}{llllll}
$a_0 = -1599.0948$ & \quad $a_1= 1064.7017$ & \quad $a_2=-262.7959$ & $a_3=31.2872$ &  $a_4=-1.8165$ &  $a_5=0.0414$ \\
$b_0 =68.2895$ & \quad $b_1=-56.3914$ & \quad $b_2=25.5238$ & $b_3=-5.3392$ &  $b_4=0.5409$ &  $b_5=-0.0216$ \\
\end{tabular}
\end{ruledtabular}
\end{table*}

\section{\label{sec:proba} Superrotor states and dissociation probability}

The laser interaction in Hamiltonian~\eqref{ham2D} is expected to drive the Cl$_2$ molecule so that the molecule acquires very high angular momentum states, and may eventually dissociate. In this way, it is natural to compute the dissociation probability as a function of the different pulse parameters, in particular, the electric field strength $F_0$ and the duration of the ramp-up $t_{\rm u}$,  the plateau $t_{\rm p}$, and the ramp-down $t_{\rm d}$. In order to understand the shape of the dissociation probability as function of the laser parameters, a method is to investigate the phase-space structures associated with Hamiltonian~\eqref{ham2D}, since the shape of the probability curves are nothing but an {\sl average} of the underlying dynamics (i.e., the phase space structures), which is largely controlled by the various pulse parameters. Because the additional half degree of freedom arising from the explicit time dependence in Hamiltonian~\eqref{ham2D} prevents the convenient exploitation of Poincar\'e sections, we do not have general-purpose tools to uncover easily the global phase-space structure of Hamiltonian~\eqref{ham2D}.  Therefore, we first analyze sample trajectories, trying to understand from that study the role of the various parameters of the laser pulse in shaping the dissociation probability curves. From an inspection of individual trajectories, we then formulate some hypotheses in the various steps of the dissociation process, and build reduce models associated with each step in this process. 

\subsection{Computation of the dissociation probability}

We compute numerically the dissociation probability $P(F_0)$ as a function of the laser amplitude $F_0$ and for different values of $t_{\rm u}$, $t_{\rm p}$ and $t_{\rm d}$. As an initial sample, we consider a large ensemble of Cl$_2$ molecules, whose dynamics is governed by the field-free Hamiltonian
\begin{equation}
\label{freeHam}
{\cal H}_0= \frac{1}{2 \mu}\left(p_r^2+\frac{p_{\phi}^2}{r^2}\right) + \varepsilon(r).
\end{equation}
Considering that initially the energy of the molecules is close to its ground state, Hamiltonian \eqref{freeHam} can be replaced by its (quantum) harmonic approximation given
\begin{equation}
\label{freeHam2}
E_0={\cal H}_0\approx -D_{\rm e}+\omega_{\rm e} \left(n+\frac{1}{2}\right) + J (J+1) B_{\rm e},\qquad n=0, 1,2, ..., \qquad J=1, 2, ...,
\end{equation}
where $\omega_{\rm e}=2.5502\times 10^{-3}~$ a.u. and $B_{\rm e}=1.1098\times 10^{-6}$ a.u..
Once the values of $n$ and $J$ are fixed, all the initial conditions $(r(0), p_r(0), \phi(0), p_{\phi}(0)=\sqrt{J(J+1)})$ of the molecules correspond to the same energy $E_0$. The values of $r(0)$ are randomly chosen in the interval $r(0)\in[r_m, r_M]$, where $r_m$ and $r_M$ are the minimum and the maximum values of $r$ allowing the condition
$$
E_0=\frac{p_{\phi}^2}{2 \mu r^2} + \varepsilon(r),
$$
to be satisfied. 
For each value of $r(0)$, the initial value value of $p_r(0)$ is given by Hamiltonian~\eqref{freeHam}. Finally, the values for $\phi(0)$ are randomly chosen in the interval $[0, 2\pi)$. Then, by the numerical integration of the equations of motion associated with Hamiltonian~\eqref{ham2D}, we propagate the ensemble of trajectories for the duration of the pulse. Typically, we consider ensembles of the size of 40000 trajectories. 
For the numerical integration of the trajectories we use the fourth order symplectic integrator BM$_6$4 of Ref.~\cite{Blanes2002} with a time step of 0.01 a.u.. The numerical codes (in Python) are available at \url{https://github.com/cchandre/OCDM}.

\begin{figure}
\includegraphics[scale=0.6]{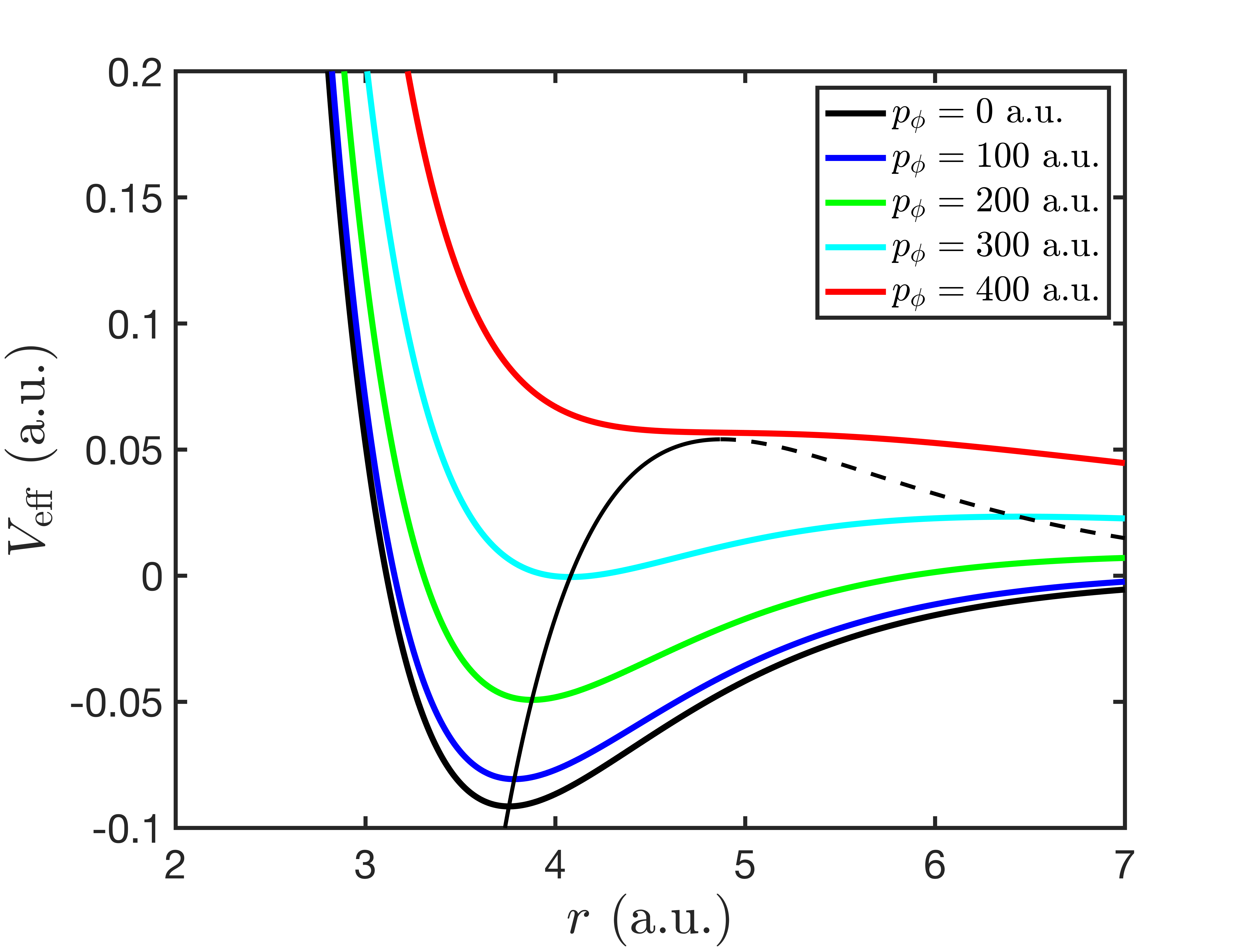}
\caption{\label{fig:Veff} Effective potential $V_{\rm eff}(r)$ for five different values of $p_\phi$. The thin continuous line indicates the location $r^*(p_\phi)$ and energy of the local minimum of the effective potential. The thin dashed line represents the location and energy of its local maximum.}
\end{figure}

In order to characterize accurately the trajectories which dissociate or not, we consider the dissociation criterion obtained from Hamiltonian~\eqref{ham2D} after the end of the laser pulse which is equivalent to considering
$$
{\cal H} = \frac{p_r^2}{2\mu}+\frac{p_\phi^2}{2\mu r^2} + \varepsilon(r).
$$
Its dynamics is equivalent to the one of a one-dimensional particle in an effective potential
$$
V_{\rm eff}(r)=\frac{p_\phi^2}{2\mu r^2} + \varepsilon(r),
$$
parameterized by $p_\phi$ which we assume to be positive in this section without loss of generality.
The effective potential is depicted in Fig.~\ref{fig:Veff} for different values of $p_\phi$. We notice the presence of a local minimum and a local maximum for small values of $p_\phi$. These minima and maxima correspond to stable and unstable equilibria, respectively. For a critical value of $p_\phi=p_\phi^*$ ($p_\phi^*\approx 394.85$ a.u.), there is a saddle-node bifurcation and there is no local minima for larger values of $p_\phi$. 

For $p_\phi>p_\phi^*$, all trajectories dissociate since there are no local extrema. For $p_\phi \leq p_\phi^*$, trajectories have the possibility to remain bounded, provided the value of $r$ is smaller than the location of the local maximum and that its energy is smaller than the height of the potential barrier. 

If initially some trajectories are in the vicinity of the minimum of the potential well around $r\approx r_{\rm e}$ and $p_\phi$ is changed adiabatically, $r(t)$ experiences larger and larger oscillations around a position $r^*(p_\phi)$ which increases from $r=r_{\rm e}$ to $r=r^*(p_\phi^*)\approx 4.87$ a.u.\ provided that $p_\phi < p_\phi^*$ (see thin black line in Fig.~\ref{fig:Veff}). For larger values of $p_\phi$, all the trajectories dissociate.  
In short, the dissociation criterion is the following one:
If the final angular momentum $p_{\phi}$ at the end of the pulse has a modulus larger than $p_{\phi}^*\approx 394.85$ a.u., dissociation takes place since for $ \vert p_{\phi} \vert >  p_{\phi}^*$, molecular bond is not possible; there is dissociation.
When $\vert p_{\phi}\vert \leq  p_{\phi}^*$, molecular bond is possible provided the final radial distance $r$ is small enough [smaller than the location of the local energy barrier of the effective potential $V_{\rm eff}(r)$] and the energy is smaller than the local maximum of the effective potential. Otherwise, the molecule dissociates.

\begin{figure}
\includegraphics[scale=0.6]{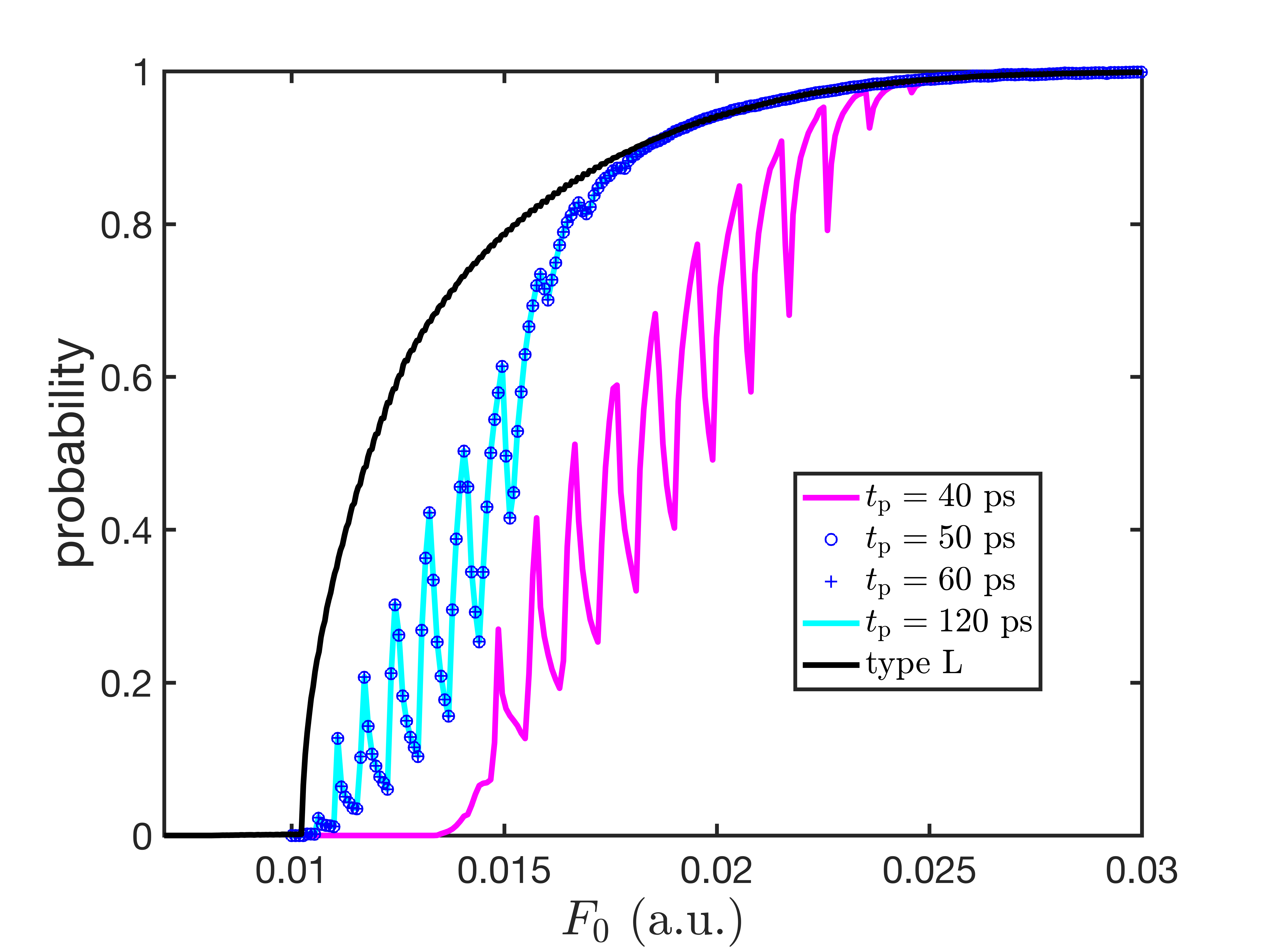}
\caption{\label{fi:proba} Dissociation probability $P(F_0)$ computed from Hamiltonian~\eqref{ham2D} for $\beta=3\times 10^{-10}$ a.u.\ and for an ensemble of initial conditions with energy given by Eq.~\eqref{freeHam2} for $n=0$ and $J=30$. The parameters of the pulse are $t_{\rm u}=t_{\rm d}=5$ ps and $t_{\rm p}=$40, 50, 60 and 120 ps. The black curve is the probability of type-L trajectories (see Sec.~\ref{sec:indtraj}).}
\end{figure}

We consider laser pulses with electric amplitude $F_0$  between $10^{-2}$ and $3\times 10^{-2}$ a.u., which corresponds to a laser fields with maximum intensity $3\times 10^{13}$ W/cm$^2$.
First, it is worth noticing that for a pulse duration of 45 ps or less, the dissociation probability is zero for all $F_0$. 
In Fig.~\ref{fi:proba} the dissociation probability $P(F_0)$  for the same ensemble of initial conditions with energy in the ground state $n=0$ and with the initial rotational state $J=30$ is represented as a function of the amplitude of the electric field $F_0$ for $\beta=3\times 10^{-10}$ a.u. and for four pulses with equal $t_{\rm u}=t_{\rm d}=5$ ps, and with $t_{\rm p}=$40, 50, 60 and 120 ps, respectively.

In the cases depicted in Fig.~\ref{fi:proba}, the dissociation probability is zero below $F_0\approx 0.013$ a.u.\ and increases sharply from 0 to 1 in the interval $0.013$ $\lesssim F_0 \lesssim 0.03$ (in atomic units), in a non monotonic way with some rather large sawtooth oscillations.
Interestingly, we find that $P(F_0)$ takes equal values for all durations of the plateau $t_{\rm p}$ larger than 50 ps.

\subsection{\label{sec:indtraj} Individual trajectories}

\begin{figure}
\includegraphics[scale=0.7]{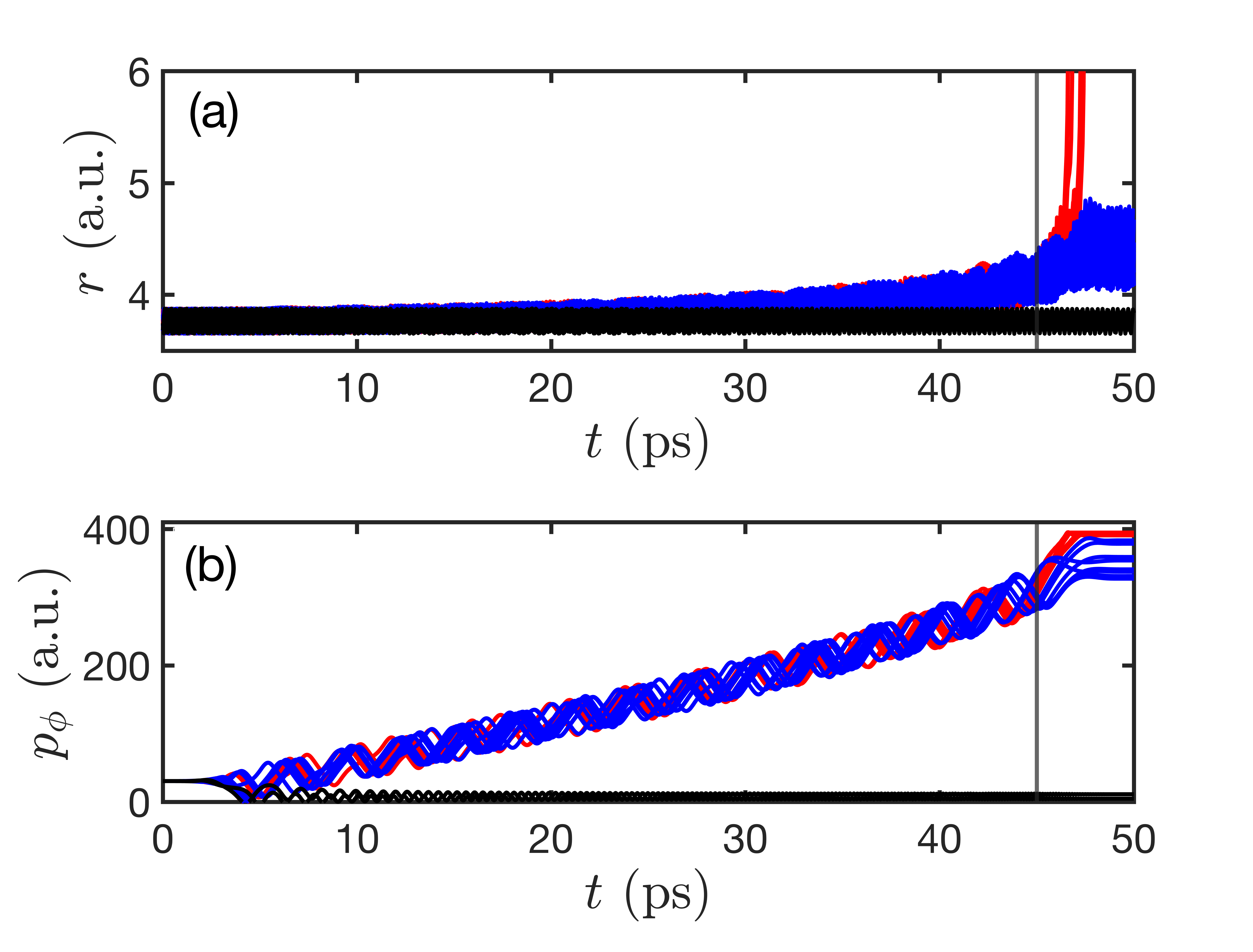}
\caption{\label{fig:trajs} Typical trajectories of Hamiltonian~\eqref{ham2D} for $F_0=0.015$ a.u. and $\beta=3\times 10^{-10}$ a.u. The upper panel displays $r(t)$, and the lower panel $p_\phi(t)$ for 20 trajectories. The parameters of the pulse are $t_{\rm u}=t_{\rm d}=5$ ps and $t_{\rm p}=40$ ps. The blue and black trajectories remain bounded at the end of the pulse, whereas the red ones are dissociating. The vertical line represents the end of the plateau of the laser pulse.}
\end{figure}
In order to analyze the dissociation probability curves $P(F_0)$, we first look at sample trajectories. In Fig.~\ref{fig:trajs}, some sample trajectories are displayed for a laser pulse with $F_0=0.015$, $\beta=3\times 10^{-10}$ a.u., $t_{\rm u}=t_{\rm d}=5$ ps and $t_{\rm p}=40$ ps. We focus on the time evolution of $r(t)$ and $p_{\phi}(t)$ of typical trajectories. From Fig.~\ref{fig:trajs}(a), we observe that, during most of the duration of the pulse, the radial distance $r(t)$ remains almost constant for all trajectories, and by the end of the pulse, some of the trajectories begin to dissociate [i.e., $r(t)$ sharply increases, red curves in Fig.~\ref{fig:trajs}(a)].
We clearly notice three main types of trajectories, and this observation is also made for other values of the parameters. We associate different colors with these three types of trajectories: The red trajectories are the ones which end up dissociating, while the black and blue remain bounded. 
From Fig.~\ref{fig:trajs}(b), roughly after the ramp-up, we notice two qualitatively different types of trajectories: the ones which experience a linear increase in angular momentum $p_\phi$ (referred to as type L), and the ones which do not (referred to as type C). The type-C trajectories are clearly not dissociating since at the end of the pulse, the radial distance $r(t)$ remains bounded. The type-C trajectories are displayed in black in Fig.~\ref{fig:trajs}.
Interestingly, we observe that, although all type-L trajectories experience a linear increase of their angular momentum during the pulse, not all the type-L trajectories eventually dissociate (only the red trajectories in Fig.~\ref{fig:trajs} do).
We remark that, regardless of their dissociation fate, all type-L trajectories end up acquiring a very large angular momentum --in a rather narrow range of values--, displaying in all cases a so-called superrotor behavior.
We remark that we have checked that all dissociating trajectories are of type L.

\begin{figure}
\includegraphics[scale=0.5]{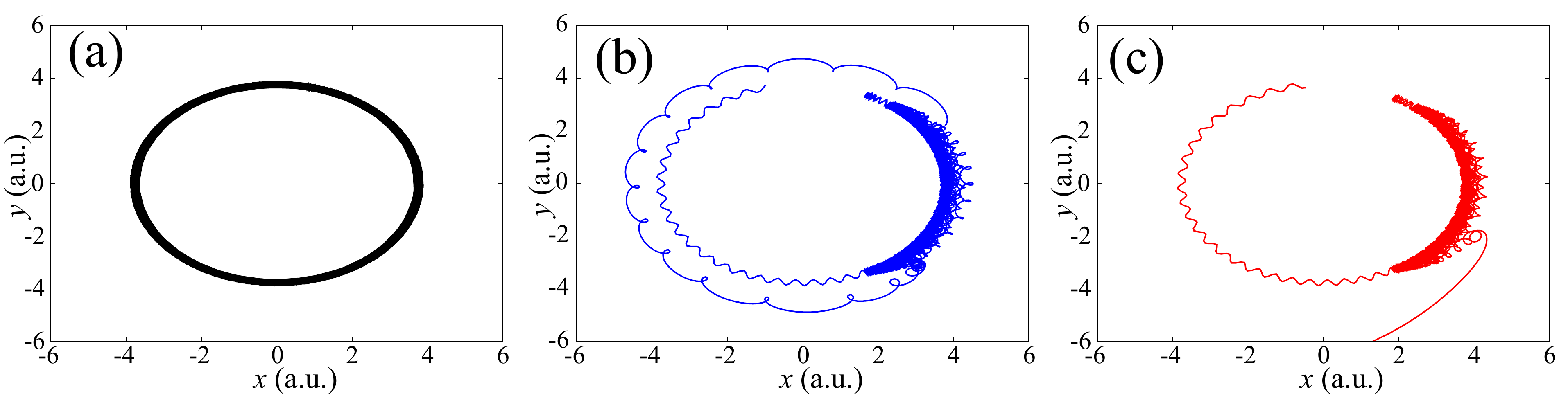}
\caption{\label{fig:trajs2} Typical trajectories of Hamiltonian~\eqref{ham2D} for $F_0=0.015$ a.u. and $\beta=3\times 10^{-10}$ a.u. (a) Trajectory of type C; (b) non-dissociating trajectory of type L; (c) dissociating trajectory of type L. The parameters of the pulse are $t_{\rm u}=t_{\rm d}=5$ ps and $t_{\rm p}=40$ ps.}
\end{figure}

It is also interesting to plot particular type-L and type-C trajectories in a Cartesian rotating frame $(x, y)$ as an alternative view of the dynamics. We find that the type-C trajectory depicted in Fig.~\ref{fig:trajs2}(a) shows a rotational behavior indicating that the field is not able to align type-C trajectories. However, the type-L trajectories in Figs.~\ref{fig:trajs2}(b)-(c) show that, after a short transient, the molecule is mostly moving in the neighborhood of the $x$ axis, which indicates a high molecular alignment along the field direction. Furthermore, whether or not they dissociate, that alignment is gained by all type-L orbits.
In Fig.~\ref{fi:proba}, we plot the occurrence probability of type-L trajectories as a function of the amplitude $F_0$ for $t_u=t_d=5$ ps and $t_p=40$ ps. We notice that this probability increases sharply and smoothly at around $F_0\approx 0.01$ a.u.\ and then saturates at around $F_0\approx 0.03$ a.u.. In particular, this probability does not display any sawtooth oscillations like the ones of the dissociation probability curves. In addition, the probability curves always remain below the occurrence probability of the type-L trajectories (which is expected since all dissociating trajectories are found to be of type L). We notice that since the angular momentum increases early in the laser pulse, the distinction between type-L and type-C is made also early in the pulse, and hence this curve is the same for an increasing duration of the plateau. 
In Fig.~\ref{fi:proba}, for large values of $F_0$ and as the length of the plateau is increased, the dissociation probability gets closer to the type-L occurrence probability. However, for intermediate values of $F_0$, the dissociation probabilities do not change significantly when the duration of the plateau increases beyond 50 ps. In other words, regardless of the value of $F_0$ between 0.01 a.u.\ and 0.017 a.u., there remains some non-dissociating type-L trajectories. 

If the field strength $F_0$ is large enough (larger than 0.017 a.u.) and the plateau is sufficiently long (longer than 50 ps), all the type-L trajectories will eventually dissociate, and the probability curve is smooth.
Otherwise, some of the type-L trajectories, although experiencing a linear increase in their angular momentum, will never dissociate, and the dissociation probability curve exhibits sawtooth oscillations.

\begin{figure}
\includegraphics[scale=0.4]{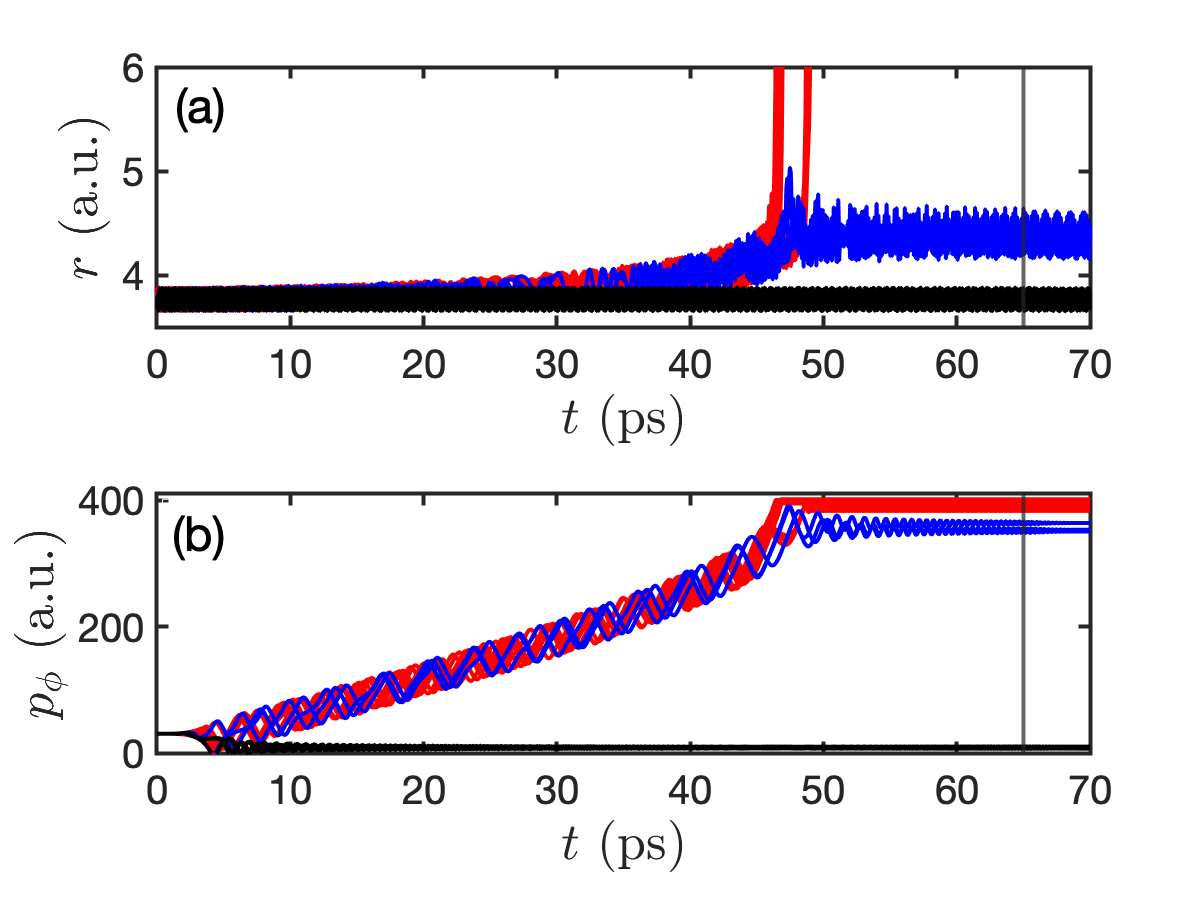}
\caption{\label{fig:trajs70} Typical trajectories of Hamiltonian~\eqref{ham2D} for $F_0=0.015$ a.u. and $\beta=3\times 10^{-10}$ a.u. The upper panel displays $r(t)$, and the lower panel $p_\phi(t)$ for 20 trajectories. The parameters of the pulse are $t_{\rm u}=t_{\rm d}=5$ ps and $t_{\rm p}=60$ ps. The blue and black trajectories remain bounded at the end of the pulse, whereas the red ones are dissociating. The vertical line represents the end of the plateau of the laser pulse.}
\end{figure}

At this stage, we rule out one possible scenario: From Fig.~\ref{fig:trajs}, we see that the dissociation or not of the type-L trajectories is happening during the ramp-down. It is tempting to attribute the dissociation fate to the ramp-down. However, the ramp-down is not responsible for this. We compute typical trajectories for a length of the plateau $t_{\rm p}=60$ ps in Fig.~\ref{fig:trajs70}. 
For a larger pulse duration, we see that there are still three types of trajectories and that the dissociation occurs at about the same time between 45 and 50 ps, so no longer during the ramp-down. 

All these observations on typical trajectories raise two main questions: What causes the linear growth of the angular momentum for some but not for all trajectories? What causes a type-L trajectory to dissociate or not? 
In order to address these questions, we investigate the nonlinear dynamics of the superrotors using reduced models. This investigation will allow us to better understand the role of the various parameters of the laser in the superrotor dynamics. 

\section{\label{sec:analysis} Nonlinear dynamics of superrotors and dissociation}

\subsection{\label{sec:angular} The angular Hamiltonian model for the superrotors}

First, we focus on the dynamical mechanism which discriminates type-L from type-C trajectories. For type-C trajectories, we notice that the distance $r$ remains close to the minimum of the potential energy surface located at $r_{\rm e}\approx 3.7560$ a.u.. After the ramp-up, type-L trajectories experience a linear growth of their angular momentum. However, during this phase of linear growth, the distance $r$ between the two atoms do not vary significantly, and remains also close to the minimum of the potential energy surface, i.e., $r\approx r_{\rm e}$.

At least during the initial phase of the pulse, a good approximation is obtained by freezing the radial degree of freedom. Hamiltonian~\eqref{ham2D} reduces to the following angular Hamiltonian
\begin{equation}
{\cal H}_{2D,{\rm red}}=\frac{p_\phi^2}{2\mu r^2}-\Omega(t) p_\phi -\frac{F_0^2}{4}\Delta\alpha(r) \cos^2\phi,
\label{eqn:angham}
\end{equation}
with one and a half degrees of freedom since $r$ is constant. In order to analyze its dynamics, we perform the following time-dependent canonical change of coordinates:
\begin{subequations}
\begin{eqnarray}
&&\widetilde{\phi}= \phi,\\
\label{eqn:canonical}
&&\widetilde{p}_\phi=p_\phi -\mu r^2 \Omega(t)=
p_\phi -\mu r^2 \beta t.
\end{eqnarray}
\end{subequations}
In this way, the explicit time dependence is removed from Hamiltonian~\eqref{eqn:angham}, and the Hamiltonian becomes
\begin{equation}
\label{eqn:angham2}
\widetilde{\cal H}_{2D,{\rm red}} = \frac{\widetilde{p}_\phi^2}{2\mu r^2}-\frac{F_0^2}{4}\Delta\alpha(r) \cos^2\widetilde{\phi}+\mu r^2 \dot\Omega(t) \widetilde{\phi}. 
\end{equation}
Hamiltonian $\widetilde{\cal H}_{2D,{\rm red}}$ is time independent since $\dot\Omega(t)=\beta$, and, therefore, it is integrable since it only has one degree of freedom. Consequently, Hamiltonian~\eqref{eqn:angham} is also integrable since the following quantity is a conserved quantity
$$
C(\phi,p_\phi,t)=\frac{(p_\phi-\mu r^2 \beta t)^2}{2\mu r^2}-\frac{F_0^2}{4}\Delta\alpha(r) \cos^2\phi +\mu r^2 \beta \phi.
$$
In order to analyze the dynamics of Hamiltonian~\eqref{eqn:angham2}, we use a dimensionless version of Hamiltonian~\eqref{eqn:angham}. Indeed, we define dimensionless time and momentum  as $\tau=t \sqrt{\beta}$ and  $p=\widetilde{p}_\phi/\mu r^2 \sqrt{\beta}$, respectively, so that Hamiltonian \eqref{eqn:angham2} becomes
\begin{equation}
\label{eqn:redHam1D}
H(x,p)=\frac{p^2}{2}+x -\eta^{-1} \cos^2 x,
\end{equation}
where $\eta=4\mu r^2\beta/(F_0^2\Delta\alpha(r))$, and the energy is measured in units of $\mu r^2 \beta$.
Here $x$ is able to take any value, not just values between $-\pi$ and $\pi$. The change of variables from the rescaled variables to the original ones is given by
\begin{subequations}
\begin{eqnarray}
\label{canonical1}
&& p_\phi = \mu r^2 \sqrt{\beta} p +\mu r^2 \beta t,\\
\label{canonical2}
&& \phi = x,
\end{eqnarray}
\end{subequations}
with an adimensional evolution parameter $\tau = t \sqrt{\beta}$ for Hamiltonian~\eqref{eqn:redHam1D}.

\begin{figure}
\includegraphics[scale=0.6]{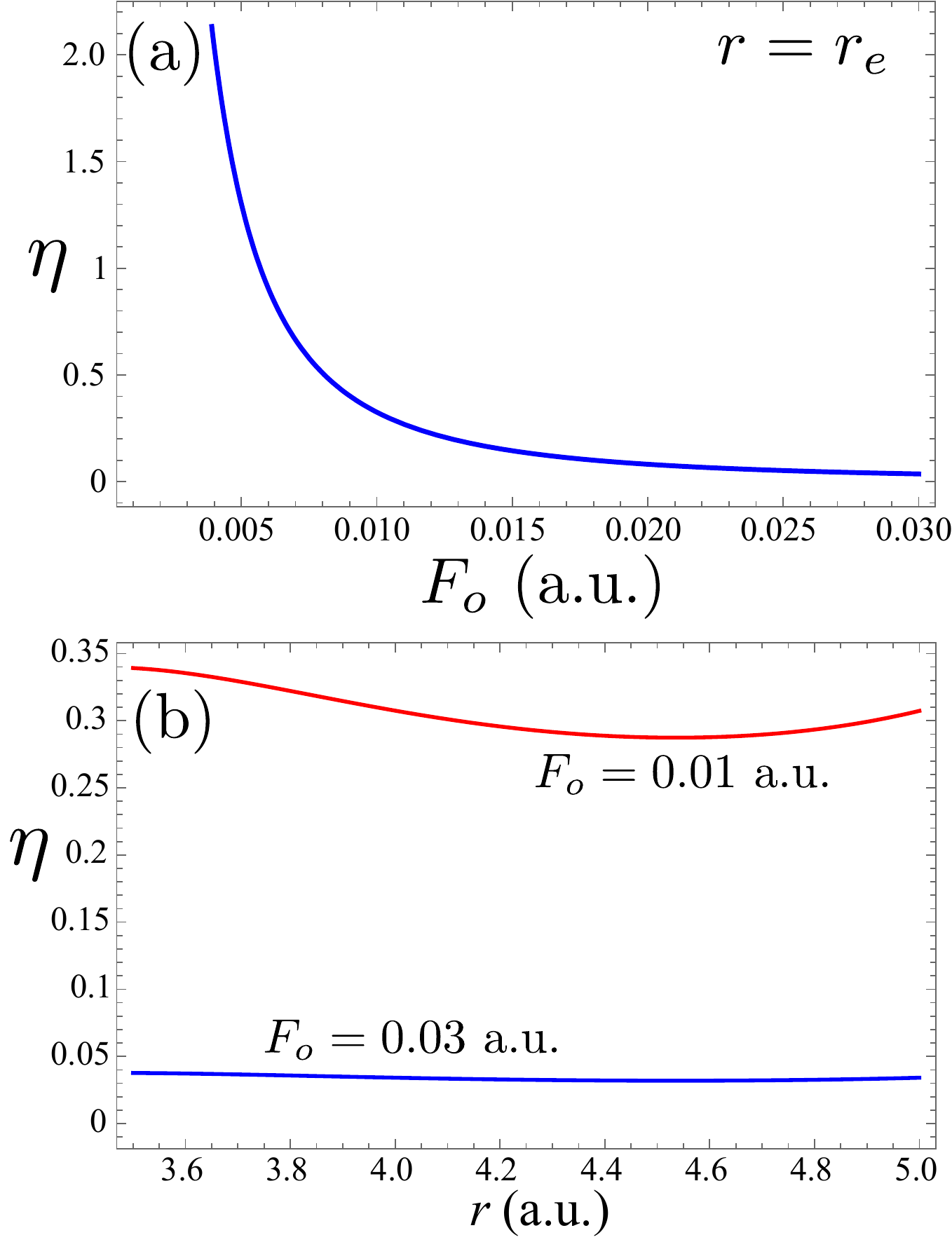}
\caption{\label{fi:eta} (a) Values of $\eta=4\mu r^2 \beta/(F_0^2 \Delta \alpha(r))$ as a function of $F_0$
at the equilibrium distance $r=r_{\rm e}$. (b) Values of $\eta$ as a function of $r$ for
$F_0=0.01$ a.u. and $F_0=0.03$ a.u. For both panels: $\beta=3\times 10^{-10}$ a.u..}
\end{figure}

This very simple one-degree-of-freedom Hamiltonian controls whether or not the molecule will experience a molecular superrotor state, i.e., when $p_\phi$ will experience a linear increase in time. The condition to have the possibility of a significant increase of the angular momentum is controlled by a single parameter, namely $\eta=4\mu r^2\beta/(F_0^2\Delta\alpha(r))$. 
For $r=r_{\rm e}$, the values of $\eta$ as a function of $F_0$ are represented in Fig.~\ref{fi:eta}(a). From Fig.~\ref{fi:eta}(b), where the values of $\eta$ as a function of $r$ for $F_0=0.01$ a.u. and $F_0=0.03$ a.u. are displayed, we infer that $\eta$ changes only slightly for varying $r$ in the neighborhood of $r=r_{\rm e}$.
The equations of motion associated with Hamiltonian~\eqref{eqn:redHam1D} are
\begin{subequations}
\begin{eqnarray}
\label{ecuMovia}
\dot x &=& p,\\
\label{ecuMovib}
\dot p &=& -1 - \eta^{-1} \sin 2 x.
\end{eqnarray}
\end{subequations}
From the equations of motion \eqref{ecuMovia}-\eqref{ecuMovib}, we deduce that, for all  $\eta \le 1$ there is an infinite number of equilibrium points given by $p^*=0$ and $x^*$ such that
\begin{equation}
\label{equiAn}
\sin 2x^*=-\eta.
\end{equation}
More explicitly, the equilibria \eqref{equiAn} are located at $x_{\rm s}(k)=-(\sin^{-1}\eta)/2 \pm k \pi$ for the stable ones and  $x_{\rm u}(k)=(\sin^{-1}\eta)/2 \pm (k+1/2) \pi$ for the unstable ones, with $k \in \mathbb{Z}$.
For a constant radial distance $r$, the equilibria \eqref{equiAn} exist if the amplitude of the electric field is sufficiently large, i.e., $F_0^2\geq 4\mu r^2 \beta/\Delta \alpha (r)$. In particular, for $r=r_{\rm e}$, those equilibria exist when $F_0 \gtrsim 0.0057$ a.u.
\begin{figure}
\includegraphics[scale=0.5]{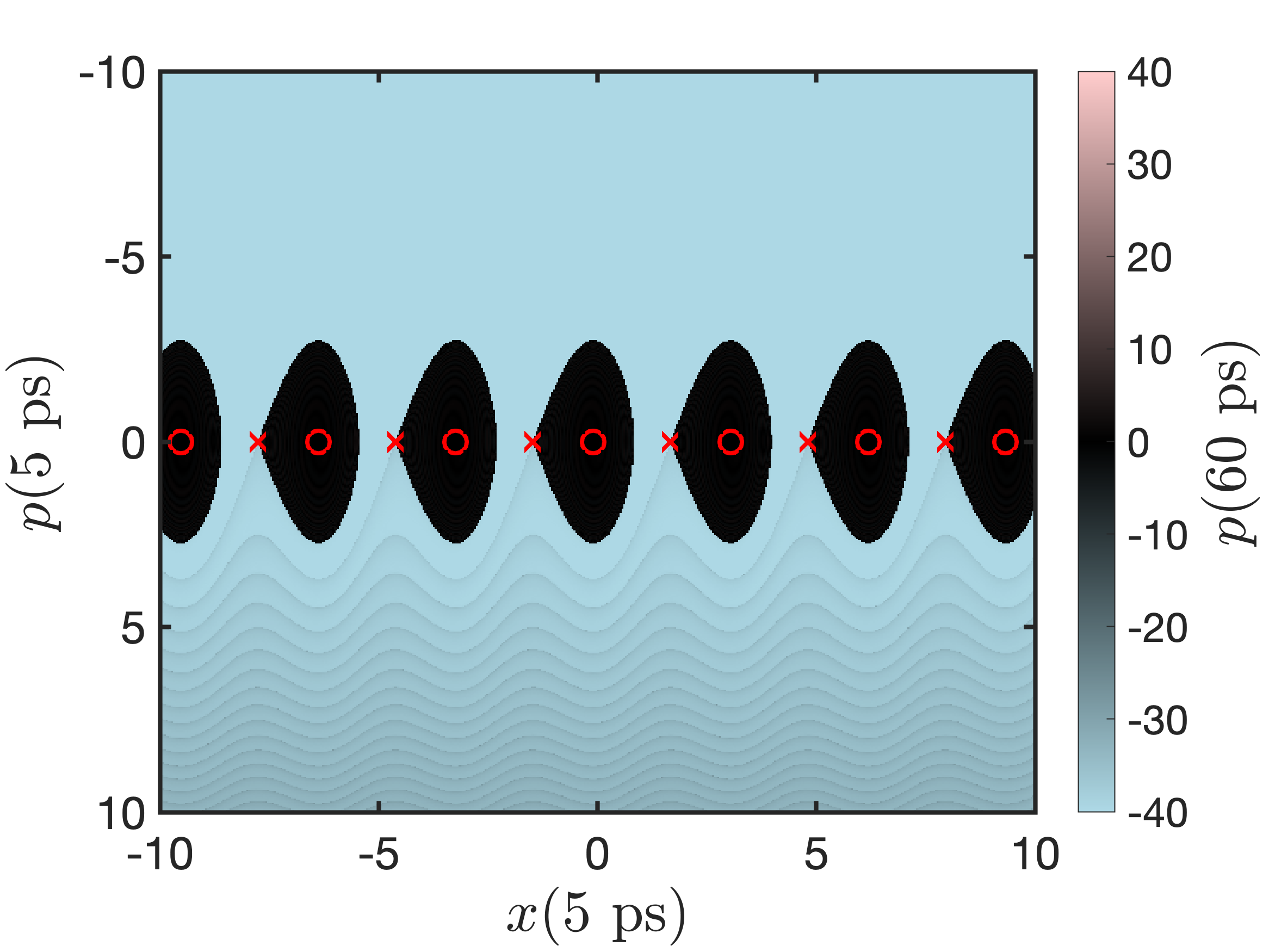}
\caption{\label{fig:phasediag} Value of the momentum $p$ at $t=60$ ps as a function of the initial conditions $(x,p)$ at the beginning of the plateau for the Hamiltonian \eqref{eqn:redHam1D} for $F_0=0.013$ a.u., $\beta=3\times 10^{-10}$ a.u.\ and $r=r_{\rm e}$. The red markers indicate some of the location of the stable ($\circ$) and unstable ($\times$) fixed points.}
\end{figure}
\begin{figure}
\includegraphics[scale=0.5]{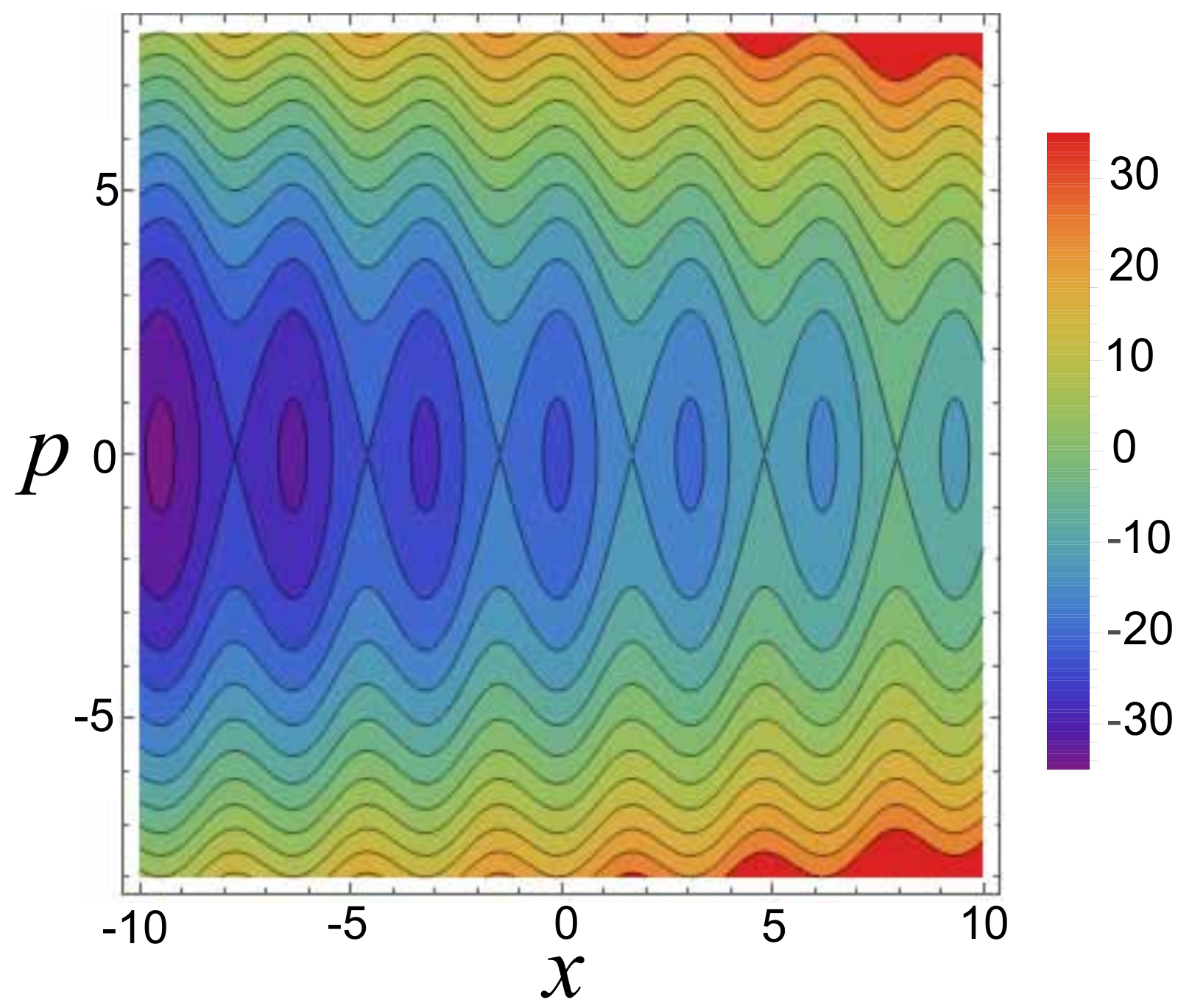}
\caption{\label{fig:phasediag2} Phase space portrait of Hamiltonian \eqref{eqn:redHam1D} for $F_0=0.013$ a.u., $\beta=3\times 10^{-10}$ a.u.\ and $r=r_{\rm e}$.}
\end{figure}
\begin{figure}
\includegraphics[scale=0.5]{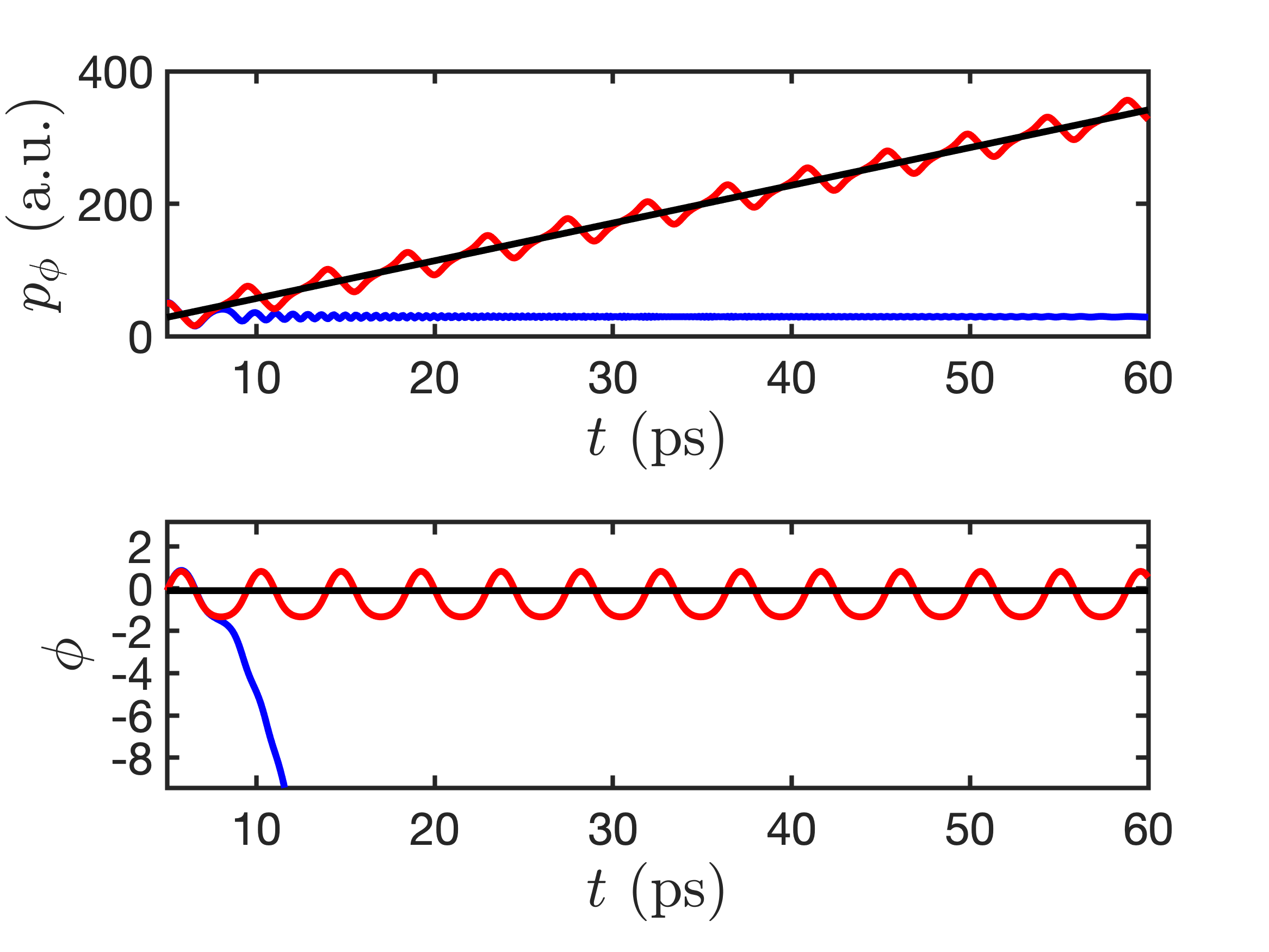}
\caption{\label{fig:traj_angular} Sample trajectories for Hamiltonian \eqref{eqn:angham2} for $F_0=0.013$ a.u., $\beta=3\times 10^{-10}$ a.u.\ and $r=r_{\rm e}$. The blue curve is for $(x,p)=(x_{\rm s}(0),2.8)$, the red curve for $(x,p)=(x_{\rm s}(0),2.7)$, and the black curve for $(x,p)=(x_{\rm s}(0),0)$.}
\end{figure}
The stable fixed points are linked to what is referred to as the ``quiet'' trajectory in Ref.~\cite{Karczmarek1999}. This quiet trajectory in the original coordinates will experience a linear increase of its momentum $p_\phi$. However, we note that this quiet trajectory or more precisely these quiet trajectories are not located at $\phi=0$ but slightly shifted by a quantity $-(\sin^{-1}\eta) /2$. In other terms, the molecule is not strictly aligned with the field as its angular momentum linearly increases. As expected, the stronger the amplitude of the field is, the more aligned the molecule will be (since $\eta$ tends to zero in this case). The explicit expression for these quiet trajectories is:
\begin{eqnarray*}
&& p_\phi(t) = \mu r^2\beta t,\\
&& \phi(t) = -\frac{1}{2}\sin^{-1}\left(\frac{4\mu r^2\beta}{F_0^2\Delta \alpha(r)} \right) \pm k \pi.
\end{eqnarray*}
The phase diagram of Hamiltonian~\eqref{eqn:redHam1D} shown in Figs.~\ref{fig:phasediag} and \ref{fig:phasediag2} displays a series of potential wells where the values of $p$ remain bounded as time increases. The center of these wells correspond to the quiet trajectory. In the neighborhood of this center, the motion is harmonic with a frequency of $(1-\eta^2)/(2\eta)$. As $\eta$ is small, this frequency is approximately $1/(2\eta)$ which means that, as the amplitude of the electric field increases, the frequency of the motion around the polarization axis is increasing (and it increases as $F_0^2$). This is the frequency of oscillations as observed in Figs.~\ref{fig:trajs} and \ref{fig:traj_angular}. 

The size of these potential wells depends on $\eta$. By looking at the isoenergetic curve corresponding to one of the unstable equilibria, we can prove that the extent in momentum of these wells is given by
$$
\Delta p = 2\sqrt{2} \left[ \frac{(1-\eta^2)^{1/2}}{\eta} -\sin^{-1}\left( (1-\eta^2)^{1/2} \right)  \right]^{1/2}.
$$
As $\eta$ increases (or equivalently as $F_0$ decreases), the size of these wells decreases since $\Delta p$ is a monotonically decreasing function of $\eta$. At $\eta=1$, there is a saddle-node bifurcation, and the extent $\Delta p$ vanishes as $(2(1-\eta))^{3/2}/3$ as $\eta$ approaches 1. As $\eta$ goes to zero, the extent $\Delta p$ diverges as $2\sqrt{2/\eta}$. It should be noted that $\Delta p$ is the extent of the range of values acquired by the type-L trajectories, and hence, related to the resolution in angular momentum of these superrotor states. For the parameters of Figs.~\ref{fig:trajs} and \ref{fig:trajs70} (where $\eta\approx 0.14$), the extent of angular momentum is approximately $\Delta p_\phi \approx 57$ a.u.. In order to increase the angular-momentum resolution, i.e., decrease $\Delta p$, the parameter $\eta = 4\mu r^2\beta/(F_0^2\Delta\alpha(r))$ needs to be increased closer to 1, e.g., through the increase of $\beta$. 

As we observe in Figs.~\ref{fig:phasediag} and \ref{fig:phasediag2}, the phase space of Hamiltonian \eqref{eqn:redHam1D} is made of regions of bounded and unbounded motions. The bounded ({\it vibrational}) phase orbits take place around the stable fixed points, while the unbounded ({\it rotational}) orbits take place between two consecutive separatrices.
Indeed, if a given initial state $(x, p)$ corresponds to one of the vibrational phase trajectories around a stable fixed point,  the change of variables \eqref{canonical1} clearly suggests that $p_\phi$ will increase linearly with time, such that the molecule will acquire a super-rotor state and, possible, will eventually dissociate. 
Conversely, if the initial state corresponds to a rotational phase trajectory, we have that, due to the increase of $p$ in time, the dynamics will be eventually dominated by the last term in Hamiltonian \eqref{eqn:redHam1D}. Therefore, for sufficiently large values of time, from the equation of motion \eqref{ecuMovib} we have that
$$
p \approx -\tau +\mbox{cte}=- \sqrt{\beta} t + \mbox{cte}.
$$
Therefore, from Eqs.~\eqref{canonical1}-\eqref{eqn:canonical} we readily obtain that $p_\phi$ will remain approximately constant in time, such that dissociation is not possible. 

The phase-space picture of the angular model provides a very good description of the mechanism by which the molecule may or may not acquire a superrotor state, i.e., why there is a clear distinction between type-L and type-C trajectories early in the pulse. Indeed, it appear that the fate of a trajectory is largely sealed at the beginning of the plateau: If the initial condition $(\phi,p_\phi)$ of a given trajectory corresponds to a state inside one of the potential wells, then that trajectory is type L because its angular momentum will increase linearly up to very large values. Otherwise, if the  initial condition $(\phi,p_\phi)$ corresponds to a state outside one of the potential wells, that trajectory is type C because the angular momentum of the molecule will remain roughly constant, and it does not dissociate (see also Ref.~\cite{Karczmarek1999}).
However, the angular model does not provide evidence explaining why some type-L orbits dissociate and others do not.

\begin{figure}
\includegraphics[scale=0.6]{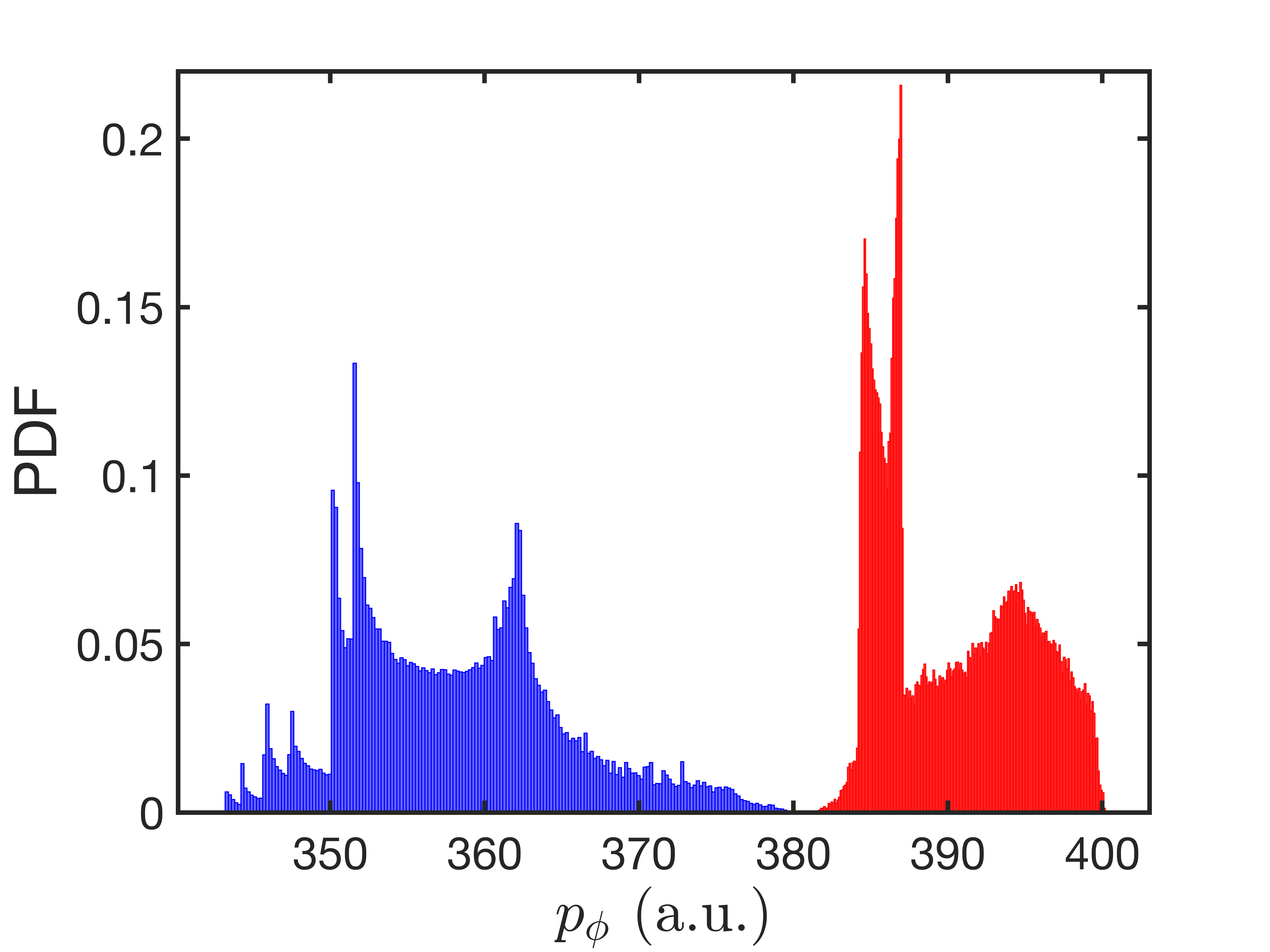}
\caption{\label{fig:dist_pphi} Probability Distribution Function (PDF) of the values of $p_\phi$ for type-L trajectories. The blue part of the histogram corresponds to non-dissociating trajectories; the red one to the dissociating trajectories. The parameters are $\beta=3\times 10^{-10}$ a.u. and $F_0=0.013$ a.u. The plateau has a duration of 160 ps. These PDFs have been computed with a sample of $10^6$ trajectories. For this sample, there are 482278 type-L non-dissociating trajectories and 169731 type-L dissociating ones.}
\end{figure}

At the end of the pulse, Fig.~\ref{fig:dist_pphi} displays the probability distribution function (PDF) of values of $p_\phi$ for type-L trajectories (see also Fig.~2 of Ref.~\cite{Karczmarek1999}). We notice that there is a clear separation of dissociating versus non-dissociating trajectories; the dissociating trajectories have larger angular momenta at the end of the pulse as expected. We could argue that this is due to the length of the plateau during which the angular momentum is increasing; this could explain that some trajectories do not have enough time to reach a critical value for dissociation. However by increasing the duration of the plateau these trajectories remain non-dissociating, ruling out this possible explanation (see Fig.~\ref{fig:trajs70}).

\subsection{The zero-velocity surface for the dissociation}

As explained in Sec.~\ref{sec:angular}, the angular model provides a good description of how superrotor type-L states are created. However, this model fails to described why some of the type-L trajectories dissociate and some do not.
To address the dissociation mechanism, we first go back to Fig.~\ref{fig:trajs}(b) [see also Fig.~\ref{fig:trajs70}(b)], where the time evolution of the angular momentum $p_{\phi}$ is displayed. In particular, we observed that type-L orbits achieve large values of $p_{\phi}$ and, interestingly, these angular momenta remain almost constant for $t\gtrsim 50$ ps. We also observe in Figs.~\ref{fig:trajs}(b) and \ref{fig:trajs70}(b) that, in all cases, the $p_{\phi}$ values of type-L orbits which end up dissociating are larger than the $p_{\phi}$ values of those type-L trajectories which remain bounded. This fact is clearly depicted in the histogram of $p_{\phi}$ in Fig.~\ref{fig:dist_pphi}.

The constant $p_{\phi}$ values attained for $t\gtrsim 50$ ps indicate that the angular model of Sec.~\ref{sec:angular} is no longer valid, such that the radial dynamics has to be taken into account. However, the constant $p_{\phi}$ values also indicate that, for $t\gtrsim 50$ ps, the $\phi$ angle would roughly behave as a cyclic variable, such that the influence of the electric field on the dynamics would be negligible. Under this assumption, for $t\gtrsim 50$ ps the dynamics will be governed by the following Hamiltonian
\begin{equation}
\label{ham2D2}
{\cal H}_{\rm 2D} \approx \frac{1}{2 \mu}\left(p_r^2+\frac{p_{\phi}^2}{r^2}\right) + \varepsilon(r)
-\Omega(t) p_{\phi}.
\end{equation}
In other words, for $t\gtrsim 50$ ps, the dynamics is mainly dominated by the centrifugal force exerted by laser field. According to Eq.~\eqref{ham2D2}, for $t\gtrsim 50$ ps the value of the angular momentum of any type-L trajectory remains almost constant.
Therefore, if at $t \approx 50$ ps the angular momentum of a given type-L orbit is larger than the critical value $p_\phi^*\approx 394.85$ a.u., such orbit will dissociate.
Conversely, if at $t \approx 50$ ps its angular momentum is smaller than $p_\phi^*$, the molecule will stay bounded since, regardless of the length of the pulse, the $p_\phi$ value of the corresponding trajectory will remain constant and therefore, below the critical value $p_\phi^*$.
\begin{figure}
\includegraphics[scale=0.6]{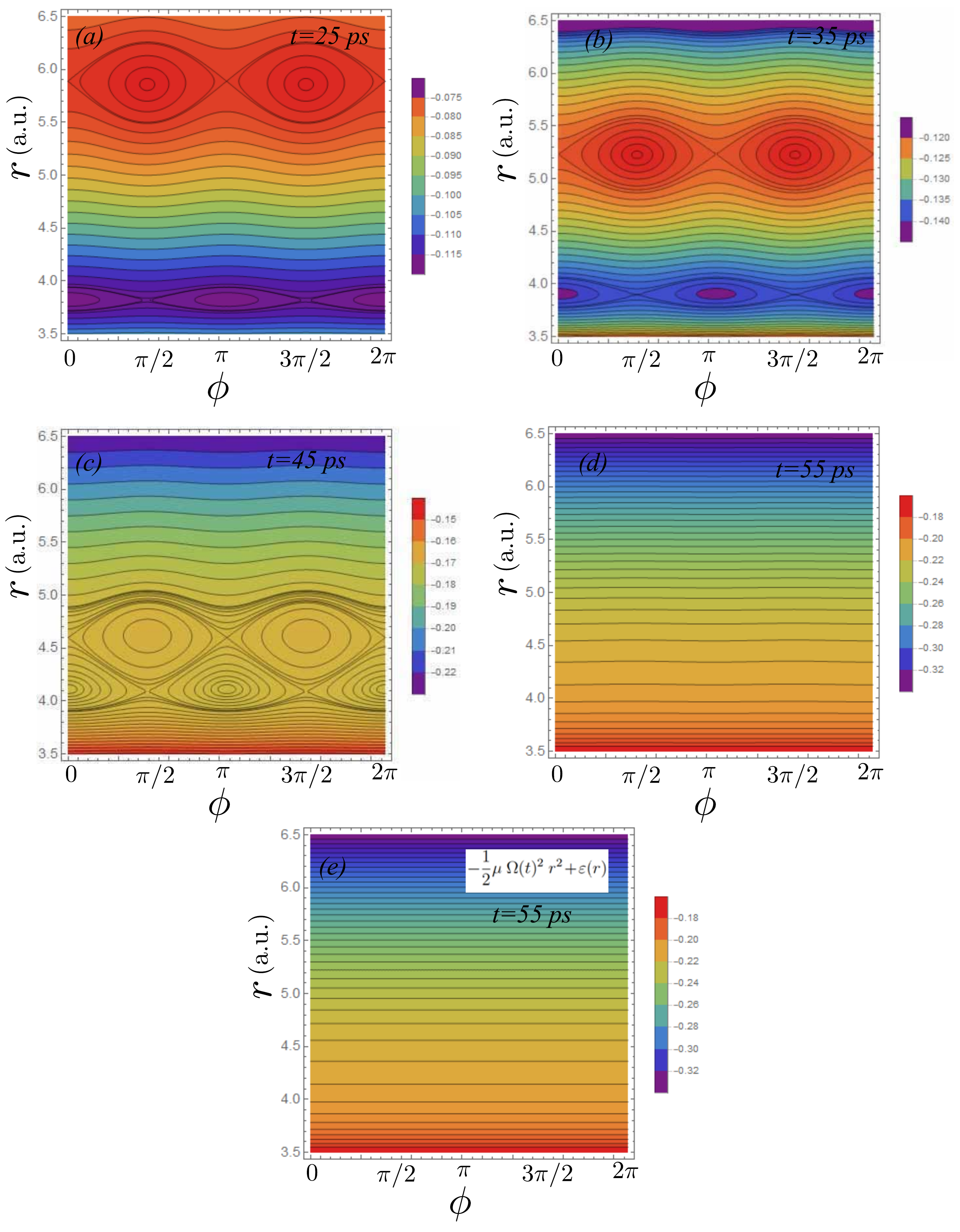}
\caption{\label{fig:figure11} Panels (a) to (d): Contour plots of the effective potential \eqref{zvs} (in atomic units).
Panel (e): contour plot of the effective potential of the Hamiltonian \eqref{ham2D2}. The parameters are $\beta=3\times 10^{-10}$ a.u., $F_0=0.015$ a.u., $t_{\rm u}=t_{\rm d}=5$ ps and $t_{\rm p}=55$ ps.}
\end{figure}
To further check the assumption that from roughly $t \gtrsim 50$ ps the dynamics is fairly described by Hamiltonian~\eqref{ham2D2}, we use the zero-velocity surface associated with Hamiltonian~\eqref{ham2D}.
The potential associated with the zero-velocity surface is defined as
\begin{equation}
\label{zvs}
U(r, \phi, t)={\cal H}_{\rm 2D}-\frac{1}{2} \mu \ (\dot x^2+\dot y^2)=-\frac{1}{2} \mu \ \Omega(t)^2 \ r^2+\varepsilon(r)
-\frac{F_0^2}{4} f(t)^2 [\Delta \alpha(r) \cos^2\phi+\alpha_\bot(r)].
\end{equation}
As it well known, the equilibrium points of the Hamiltonian flow are the critical points of the effective potential $U(r, \phi, t)$.
For a given value of the field $F_0$, we plot stroboscopically the effective potential $U(r, \phi,t)$ at different times. For example, in Fig.~\ref{fig:figure11} a typical stroboscopic evolution of the contour maps of  $U(r, \phi,t)$ for $F_0=0.015$ a.u. and $t=$25, 35, 45 and 55 ps is shown. The pulse parameters are  $t_u=t_d=5$ ps and $t_p=50$ ps.
At $t=25$ ps we observe in Fig.~\ref{fig:figure11}(a) that $U(r, \phi,t)$ has two equivalent minima at $\phi=0, \ \pi$ and $r\approx 3.8230$ a.u. The potential wells around these minima are separated by a separatrix passing through two saddle points located at $\phi=\pi/2, \ 3\pi/2$ and $r\approx 3.8230$ a.u.. At the larger value $r\approx 5.8990$ a.u.\ and at $\phi=\pi/2, \ 3\pi/2$ there are two equivalent maxima separated by a separatrix passing through two saddle points located at $\phi=0, \ \pi$ and $r\approx 5.8990$ a.u.. 
At $t=35$ ps, we observe in Fig.~\ref{fig:figure11}(b) that the two maxima approach the minima, remaining the later almost at the same position. At $t=45$ ps, the maxima and the minima regions are very close each other (see Fig.~\ref{fig:figure11}(c)), such that for $t\approx 47$ ps, the minima and the saddle points at $\phi=0, \ \pi$ and the maxima and the saddle points at $\phi=\pi/2, \ 3\pi/2$ collide. For $t\gtrsim 47$ ps, the effective potential  $U(r, \phi,t)$ does not present critical points (see Fig.~\ref{fig:figure11}(d) for $t=55$ ps). Because longer times means larger values of the rotating frequency $\Omega(t)$, the time evolution of $U(r, \phi,t)$ depicted in Fig.~\ref{fig:figure11} indicates that, for $t\gtrsim 47$ ps, most of the dynamics is governed by the rotating term $\Omega(t) p_{\phi}$ in Hamiltonian~\eqref{ham2D}. Furthermore, there are no significant differences between the contour plot in Fig.~\ref{fig:figure11}(d), and the contour plot of the effective potential of Hamiltonian~\eqref{ham2D2} which is simply $\varepsilon(r) -\mu \Omega(t)^2 r^2/2$ (see Fig.~\ref{fig:figure11}(e)).

The expression of the approximate time at which the bifurcation occurs is given by
\begin{equation}
\label{eqn:tb}
t_{\rm b}=\frac{1}{\beta}\left(\frac{\gamma D_{\rm e}}{2\mu \left(r_{\rm e} +\frac{\log 2}{\gamma} \right)} \right)^{1/2}.
\end{equation}
This approximation is obtained by looking at the condition under which $\varepsilon(r) -\mu\Omega(t)^2 r^2/2$ has local extrema since the contribution proportional to $F_0^2$ in Eq.~\eqref{zvs} is small. The time and location of these extrema are linked by the condition $\beta^2 t^2=\varepsilon^\prime(r)/(\mu r)$. By investigating the behaviour of the function $\varepsilon^\prime(r)/r$, we approximate the location of its maximum at $r\approx r_{\rm e}+\log 2 / \gamma$. Combining these elements, we establish the condition~\eqref{eqn:tb}.
For the chosen parameters, we have $t_{\rm b}\approx 47.5$ ps.  
For $t\geq t_{\rm b}$, there is no longer a minimum of the potential well $U$ and, therefore, the motion can potentially become unbounded, depending on the value of its angular momentum at this specific time. 

\section*{Conclusion}

In this article, we investigated the mechanisms and the conditions under which a diatomic molecule in an optical centrifuge acquires superrotor states which can potentially lead to dissociation.
To carry out this study, we considered the chlorine molecule Cl$_2$ as an example. In addition to the molecular potential energy curve, our Hamiltonian model includes accurate radial functions for the parallel and perpendicular components of the molecular polarizability through which the interaction between the laser field of the optical centrifuge and the molecule takes place.

The mechanisms and the conditions under which superrotor states are created and potentially lead to dissociation of the molecule have been investigated as functions of the parameters of the laser, namely, its amplitude, the duration of the laser pulse, and the acceleration of the rotation of the polarization axis. The angular degree of freedom is responsible for the creating of superrotor type-L trajectory. The  condition for the existence of potential type-L states is given by $4\mu r_{\rm e}^2\beta/(F_0^2 \Delta\alpha(r_{\rm e}))\leq 1$. Under this condition, stable equilibria ensure the quasi-alignment of the molecular axis with the polarization axis under specific initial conditions (inside potential wells), and hence a linear increase of the angular momentum. The radial degree of freedom is mostly responsible for the dissociation dynamics: The local minima and maxima of the zero-velocity surface ensure that the interatomic distance $r$ remains bounded until these extrema collide and $r$ is potentially unbounded. We have estimated this critical time $t_{\rm b}$ to be given by Eq.~\eqref{eqn:tb}.  The analysis of the nonlinear dynamics provides a way to control the different states at the end of the laser pulse by adjusting the parameters of the laser field.  
For instance, in order to suppress dissociation and have all type-L trajectories bounded, the duration of the pulse has to be shorter than $t_{\rm b}$, i.e.,
$$
t_{\rm u}+t_{\rm p} +t_{\rm d} \leq t_{\rm b}.
$$

\begin{acknowledgments}
JPS acknowledges financial support by the Spanish Project No. MTM2017-88137-C2-2-P
(MINECO), and the hospitality of the Institut Fresnel during his stay at Aix-Marseille Universit\'e.
\end{acknowledgments}

\section*{Author contributions}

{\bf C. Chandre}: Conceptualization (equal); Formal analysis (equal);
Funding acquisition (equal); Investigation (equal); Methodology (equal);
Software (equal); Writing – original draft (equal). {\bf J.P. Salas}: Conceptualization (equal); Formal analysis (equal);
Funding acquisition (equal); Investigation (equal); Methodology (equal);
Software (equal); Writing – original draft (equal).

\section*{Data availability}

Data sharing is not applicable to this article as no new data were created or analyzed in this study.


%

\end{document}